\documentclass[aps,prd,preprint,groupedaddress]{revtex4-1}

\usepackage{graphicx}
\usepackage[table,xcdraw]{xcolor}
\usepackage{xcolor}
\usepackage{amsmath}
\begin{document}

\title{ Configuration entropy description of charmonium  dissociation under the influence of magnetic fields   }

\author{Nelson R. F. Braga}\email{braga@if.ufrj.br}
\author{Rodrigo da Mata}\email{rsilva@if.ufrj.br }

\affiliation{Instituto de F\'{\i}sica,
Universidade Federal do Rio de Janeiro, Caixa Postal 68528, RJ
21941-972 -- Brazil}


\begin{abstract} 
Heavy ion collisions, produced in particle accelerators,  lead to the formation 
of a new state of matter,  known as the quark gluon plasma. It is not possible to observe directly the plasma, where quarks and gluons are not confined into  hadrons. All the available information comes from the particles that reach the detectors after the strongly interacting matter hadronizes. Among those particles, one that plays an important  role is the charmonium      J/$\psi$ 
heavy meson, made of a  $ c  \bar c$  quark anti-quark pair. The fraction of such particles produced in a heavy ion collision is related to the  dissociation level caused by  the plasma. On the other hand, the dissociation of $J/\Psi  $ in the plasma is influenced by the temperature and  the density of the medium and also by the presence of magnetic fields, that  are produced in non central collisions. 

A very interesting tool to study stability of physical systems is the configuration entropy (CE). 
In recent years many examples in various kinds of physical systems appeared in the literature, where an increase in the CE is associated with an increase in the instability of the system. 
In this article we calculate the CE for charmonium quasistates inside a plasma with a magnetic field background, in order to investigate how the instability, corresponding in this case to the   dissociation in the thermal medium,  is translated into the dependence of the CE on the field.

\end{abstract}


\maketitle

 \section{ Introduction }   
  
 In the recent years many examples examples appeared in the literature, where the configuration 
entropy (CE)  \cite{Gleiser:2011di,Gleiser:2012tu,Gleiser:2013mga}  provides information about  stability  of physical systems. It was found, for diverse systems, as for example 
\cite{Gleiser:2015rwa,Bernardini:2016hvx,Bernardini:2016qit,Braga:2017fsb,Braga:2018fyc,Bernardini:2018uuy,Ferreira:2019inu,Alves:2014ksa,Correa:2015vka,Sowinski:2017hdw,Casadio:2016aum,Correa:2016pgr,Braga:2016wzx,Karapetyan:2016fai,Karapetyan:2017edu,Alves:2017ljt,Karapetyan:2018oye,Karapetyan:2018yhm,Gleiser:2018kbq,Colangelo:2018mrt,Lee:2018zmp,Bazeia:2018uyg,Ma:2018wtw,Zhao:2019xle,Ferreira:2019nkz,Braga:2019jqg,Bernardini:2019stn,Karapetyan:2019ran,Braga:2020myi,Karapetyan:2020yhs,Ferreira:2020iry,Alves:2020cmr,MarinhoRodrigues:2020ssq},    that the more stable, the lower is the value of the CE. 
 
Instability may be a consequence of many different types of transitions, depending on the system one considers. For the particular case that we study here  --  charmonium quasistates inside a plasma --  the transition corresponds to the thermal dissociation in the  medium. 
A natural question to be asked is: why is it important to study the stability of charmonium in a thermal medium?  The point is that heavy vector mesons, like charmonium, that is formed by a  $ c \bar c $ valence  quark anti-quark pair may survive the dissociation process that affects the ligth hadrons, when the quark gluon plasma (QGP) is formed in a heavy ion collision. 
So, the fraction of charmonium detected after a heavy ion collision can be used as a source of information about the QGP  \cite{Matsui:1986dk,Satz:2005hx}. 
Interesting reviews about the QGP  are found in, for example: \cite{Bass:1998vz,Scherer:1999qq,Shuryak:2008eq,CasalderreySolana:2011us}. 

The dissociation of charmonium in the plasma is affected by the  temperature, by the  density   and also by the presence of background magnetic fields.
It is possible to describe such a behavior using a holographic bottom up model  \cite{Braga:2015jca,Braga:2016wkm,Braga:2017oqw,Braga:2017bml,Braga:2018zlu,Braga:2018hjt,Braga:2019yeh,Braga:2019xwl}. In this article we are concerned with the effect of the magnetic field on the charmonium quasistates inside a plasma and the corresponding instability caused by the dissociation in the medium. 

The basis for the definition of the configuration entropy is the Shannon information entropy  \cite{shannon}  that represents the information contained in a variable $x$  that assumes discrete values $x_n$ with  probabilities $p_n$:
\begin{equation}
  - \sum_n  \, p_n  \log p_n \,.
 \label{discretepositionentropy}
 \end{equation}
  In order to introduce the   configuration entropy \cite{Gleiser:2013mga}  one takes a normalizable function in coordinate space $ \rho ({\vec r}) $ (in general the energy density of a physical system) 
and transforms to momentum space: 
  \begin{equation}
 {\tilde \rho  } (\vec k )  = \frac{1}{(2\pi)^{d/2}} \int d^dr \,  \rho ({\vec r}) \exp ( -i \vec k \cdot \vec r )   \,. 
 \label{Conjugate}
 \end{equation}
 Then one defines the modal fraction:  
 \begin{equation}
  {\tilde  \epsilon } ({\vec k}) =\frac{  \vert {\tilde \rho } ({\vec k}) \vert^2 }{  \int d^dk \, \vert {\tilde \rho } ({\vec k}) \vert^2}  \,,
  \label{MMF}
  \end{equation}
and the  CE is introduced as:
 \begin{equation}
 \tilde{S} = - \int d^dk \, { \tilde  \epsilon} ({\vec k}) \log  {\tilde  \epsilon} ({\vec k}) \,.
 \label{momentumentropy}
 \end{equation}
 In contrast to the discrete case, for continuum variables  this quantity may  be negative. 
 In this case, one can alternatively \cite {Gleiser:2018kbq}  define a different type of modal fraction
  \begin{equation}
    \epsilon   ({\vec k}) =\frac{  \vert {\tilde \rho } ({\vec k}) \vert^2 }{   \vert {\tilde \rho } ({\vec k}) \vert_{max}^2 }  \,,
  \label{DMF}
  \end{equation}
    where  in contrast to eq.  (\ref{MMF})  we do not normalize the function  $  \tilde \rho   ({\vec k})   $
    but rather divide by the maximum value of the square of the absolute value:
    $ \vert {\tilde \rho } ({\vec k}) \vert_{max}^2 $  . Then one introduces the so called differential configuration entropy (DCE):
  \begin{equation}
  S = - \int d^dk \,  \epsilon  ({\vec k}) \log  { \epsilon } ({\vec k}) \,.
 \label{Dmomentumentropy}
 \end{equation}
 We calculate in this article the DCE for charmonium inside a plasma in the presence of magnetic fields using a holographic model to describe the quasistates in the medium. 
  
  This article is organised in the following way: in section II   we review the holographic model for charmonium in a plasma in the presence of magnetic fields. Then in section III we develop  the calculation of the configuration entropy of the  J/$\psi$  quasistates inside the plasma. In section IV we present the results and section V contais some final conclusions.

\section{Holographic description of charmonium in a plasma  with magnetic fields }

Charmonium J$/\psi $ vector mesons inside a plasma with magnetic field can be studied using the holographic model 
 of  Ref. \cite{Braga:2019yeh}.   They are represented by a 5-dimensional  dual vector field 
  $V_m  $   with an action integral  of the form
     
\begin{equation}
I \,=\, \int d^4x dz \, \sqrt{-g} \,\, e^{- \phi (z)  } \, \left\{  - \frac{1}{4 g_5^2} F^{\ast}_{mn} F^{mn}
\,  \right\} \,\,, 
\label{vectorfieldaction}
\end{equation}
with  $F_{mn} = \partial_m V_n - \partial_n V_m$. 

In the absence of the plasma and of magnetic fields, that means, in the vacuum, the space is just a five dimensional anti-de Sitter  one. In this case the masses and decay constants of  charmonium states are obtained from the background  field $\phi(z)$ :  
\begin{equation}
\phi(z)=k^2z^2+Mz+\tanh\left(\frac{1}{Mz}-\frac{k}{ \sqrt{\Gamma}}\right)\,.
\label{dilatonModi}
\end{equation} 

 The   three energy parameters introduced in the model are interpreted as:   $k$ represents the  quark mass,   $\Gamma $  the string tension of the quark anti-quark interaction
 and $M$ is a large mass associated with the charmonium non-hadronic decay,  when the heavy meson decays into leptons, that  involves the matrix element  $ \langle 0 \vert \, J_\mu (0)  \,  \vert    \textrm{ J}  /\psi  \rangle = \epsilon_\mu f_n m_n \,$.   The  values that provide the best fit to the spectrum are \cite{Braga:2018zlu}:
\begin{equation}
  k_c  = 1.2 \, \textrm{GeV } \, , \,   \sqrt{\Gamma_c } = 0.55 \, \textrm{GeV }  \,   , \,\, M_c=2.2 \, \textrm{GeV }  \, .
  \label{parameters}
  \end{equation}     
  
  The extension to finite temperature and in the presence of a constant magnetic field $eB$, pointing in the $x_3$ direction,   is obtained using in the action of eq. (\ref{vectorfieldaction})  the same scalar field background  (\ref{dilatonModi})  of the vaccum case but with  the following black hole  geometry\cite{Dudal:2015wfn} :
  
 \begin{equation}  
 ds^2 \,\,= \,\, \frac{R^2}{z^2}  \,  \Big(  -  f(z) dt^2 + \frac{dz^2}{f(z) }  +( dx_1^2+dx_2^2)d(z)+dx_3^2h(z)  \Big)   \,,
 \label{metric3}
\end{equation}
 where
\begin{eqnarray}
f (z) &=& 1 - \frac{z^ 4}{z_h^4}+\frac{2}{3}\frac{e^2B^2z^4}{1.6^2}\ln\left(\frac{z}{z_h}\right)\, ,
\\
h(z) &=& 1 + \frac{8}{3}\frac{e^2B^2}{1.6^2}\int^{1/z}_{+\infty}dx\frac{\ln{(z_h x)}}{x^3(x^2-\frac{1}{z_h^4x^2})}\,, \\
d(z) &=& 1 - \frac{4}{3}\frac{e^2B^2}{1.6^2}\int^{1/z}_{+\infty}dx\frac{\ln{(z_h x)}}{x^3(x^2-\frac{1}{z_h^4x^2})}\, .
\end{eqnarray}
 
The plasma temperature is given by: 
\begin{equation} 
T =  \frac{\vert  f'(z)\vert_{(z=z_h)}}{4 \pi  } = \frac{1}{4\pi}\left\vert \frac{4}{z_h}-\frac{2}{3}\frac{ e^2B^2z_{h}^3} { 1.6^2} \right\vert \,.
\label{temp}
\end{equation} 
  
  One can find interesting alternative holographic studies of  heavy flavour hadrons, for example,  in  
 \cite{Branz:2010ub,Gutsche:2012ez,Fadafan:2011gm,Fadafan:2012qy,Fadafan:2013coa,Afonin:2013npa,Mamani:2013ssa,Liu:2016iqo,Liu:2016urz,Ballon-Bayona:2017bwk,Dudal:2018rki,Gutsche:2019blp,Bohra:2019ebj,Zhang:2019qhm,MartinContreras:2019kah}.     Also,  magnetic field effects in  hadronic matter were anlysed before in many references, like  
 \cite{Ballon-Bayona:2013cta,Mamo:2015dea,Evans:2016jzo,Li:2016gfn,Ballon-Bayona:2017dvv,Rodrigues:2017cha,Iwasaki:2018pby,Giataganas:2018uuw,Iwasaki:2018czv}.

 \section{Configuration entropy of charmonium in a medium with magnetic fields } 
 
 Now we follow the necessary steps in order to calculate the differential configuration entropy, defined in eq. (\ref{Dmomentumentropy}), for the case of charmonium in a plasma with magnetic fields. The quantity to be Fourier transformed as in eq. (\ref{Conjugate}) and then used to calculate de modal fraction using eq. (\ref {DMF}) is the energy density of the charmonium   J/$\psi$  quasistate, that corresponds to the $ T_{00} $ component of the energy momentum tensor. In order to represent a meson at rest one considers a solution for the vector field of the form: $V_{\mu}=\eta_{\mu}v(\omega , z)e^{-i \omega t}$ and choose the radial gauge $V_z=0$. 
 So, the energy density depends only on the coordinate $z$ of the charged AdS Black hole metric of 
 eq. (\ref{metric3}).  We assume that in our effective model, described by the action integral
 of eq. (\ref {vectorfieldaction}), the energy momentum tensor $ T_{mn} $ can be obtained  as in general relativity:
\begin{equation}\label{EnTe}
T_{mn}(z)=\frac{2}{\sqrt{-g}}\!\left[ \frac{\partial(\sqrt{-g}\mathcal{L})}{\partial g^{mn}}-\frac{\partial}{\partial x^{p}}\frac{\partial(\sqrt{-g}\mathcal{L})}{\partial\left(\frac{\partial g^{mn}}{\partial x^{p}} \right)} \right] \,. 
\end{equation}
 For the vector field action   (\ref {vectorfieldaction})  one finds  
\begin{equation}
\!\!\!\!\!\! \rho (z) \,=\, T_{00} (z) \!=\!\frac{e^{- \phi (z)  } }{g^{2}_{5}}\!\left[g_{00}\!\left(\frac{1}{4}g^{mp}g^{nq}F_{mn}F_{pq}\right)\!-\!g^{mn}F_{0n}F_{0m}\right]\,.
\end{equation}   

The magnetic field is pointing in the $x_3$ direction. Its is convenient to separate the analysis in two different cases. The transverse  one, when the polarisation is perpendicular to the magnetic field, corresponding to the vector field in the direction of $\eta_{\mu_{ T}}=(0,cos(\alpha),sin(\alpha),0)$ and a longitudinal one, in the direction of $\eta_{\mu_{L}}=(0,0,0,1)$.
 
 For the transversal  case one finds  the energy density 
 
\begin{equation}\label{rhoT}
 \!\!\!\!\!\!\rho_{T} (z)\!=\!\frac{z^2 e^{- \phi (z)  } }{2 R^2 g^{2}_{5}}\!\left(\frac{1}{d(z)}\right)\left[\frac{f^{2}(z)}{|\omega|^{2}} |E'_{\alpha}|^2 +|E_{\alpha}|\right]\,,
\end{equation} 
where we introduced the electric field   defined as $E_{\alpha}= \omega V_{\alpha}$ as usual.  For the longitudinal case the density reads
\begin{equation}\label{rhoL}
 \!\!\!\!\!\!\rho_{L} (z)\!=\!\frac{z^2 e^{- \phi (z)  } }{2 R^2 g^{2}_{5}}\!\left(\frac{1}{h(z)}\right)\left[\frac{f^{2}(z)}{|\omega|^{2}} |E'_{3}|^2 +|E_{3}|\right]\,, 
\end{equation}
where, in a similar way $E_{3}= \omega V_{3}$.  

Our next task is to find  the   solutions of the vector field equations of motion that represent the quasistates of charmonium. These solutions, called quasinormal modes,  must satisfy 
 infalling boundary conditions at the  event horizon  $z = z_h $   and  must also  vanish at the boundary $ z = 0$ in order to ensure normalizability. This pair of conditions is in general satisfied 
 by complex frequencies    $\omega $ with a real component associated with  the thermal mass and an imaginary component associated with the thermal width of the quasistate.  
 
Writing the  equations of motion obtained from the action (\ref{vectorfieldaction})  with  metric    (\ref{metric3})  in terms of  the electric fields  one finds 
\begin{equation}
\label{eqT}
E_{\alpha}''+\left(\frac{f'(z)}{f(z)}-\frac{1}{z}-\phi'(z)+ \frac{h'(z)}{2h(z)}\right)E_{\alpha}'+\frac{\omega^{2}}{f^2} E_{\alpha} = 0\,,  \,\,\, (\alpha = 1,2) \,, 
\end{equation} 

\begin{equation}
\label{eqL}
E_{3}''+\left(\frac{f'(z)}{f(z)}-\frac{1}{z}-\phi'(z)+\frac{d'(z)}{d(z)}-\frac{h'(z)}{2h(z)}\right)E_{3}'+\frac{\omega^{2}}{f^2} E_{3} = 0\,, 
\end{equation} 
where  the prime means derivative with respect to  $z$.

In order to impose the infalling boundary conditions on the horizon it is helpful to write the equations of motion in terms of the tortoise coordinate $r_{*}$ defined by $\partial_{r_{*}} = - f(z)\partial_{z}$ with $r_{*}(0)=0$,  for z in the $0 \leq z \leq zh$. In terms of  $r_{*}$ 
 the solutions split, near the horizon,  into a combination of   infalling  and outgoing waves.  Both  equations (\ref{eqT}) and (\ref{eqL}) can be written in the generic  form:
\begin{equation}
\label{eqgeral}
E''+ a(z) E'+ b(z) E = 0\,, 
\end{equation} 
where the  functions $a(z)$ and $b(z)$ are different for each polarisation. One searches for a  field redefinition of the form $\psi = e^{-\frac{\zeta(z)}{2}}E$  such that equation   (\ref{eqgeral}), written   in terms of the new function $\psi $ and of the  tortoise coordinate,  takes the form of  a  wave equation

\begin{equation}
\label{Sequation}
\partial^{2}_{r_*}\psi +\omega^{2}\psi = U\psi \,,
\end{equation}
as long as the  derivative of the  function $\zeta(z)$ satisfies 
\begin{equation}
\zeta'(z) = \frac{f'(z)}{f(z)}-a(z) \,. 
\end{equation}  
The potential $U(z)$ has the form 
\begin{equation}
\label{potencial}
U(z) = -f^2(z)\left\{\left(\frac{\zeta'(z)}{2}\right)+\frac{\zeta''(z)}{2}+a(z)\frac{\zeta'(z)}{2}\right\}\,.
\end{equation}
For both transversal and longitudinal polarisations the potential   diverges  at $z=0$. So, the normalizability  condition for the quasinormal mode solutions require that one   must impose the boundary condition $\psi(z=0)=0 $. At the horizon the potential vanishes:  $U(z=z_h)=0$. 
So, in the limit $z \to z_h$  the general solution is a combination of   infalling    $\psi = e^{-i\omega r_*}$ and  outgoing $\psi = e^{+i\omega r_*}$ ones. 
 The relevant incoming solutions can be expanded near the horizon in the form
\begin{equation}
\psi=e^{-i\omega r_*(z)} \left[1+c^{(1)}\left(z-z_h\right)+\dots\right] \,.
\label{expansion}
\end{equation}
The potential can also be expanded near the horizon as:
\begin{equation}
U=(z-z_h)U'(z_h)+\dots\,.
\label{Uexpansion}
\end{equation}
The coefficient $c^{(1)}$ in eq. (\ref{expansion})  is given by
\begin{equation}
c^{(1)}=\frac{U'(z_h)}{f'^{2}(z_h)+2 i \omega f'(z_h)},.
\label{coeficiente}
\end{equation}
Following this approach, que boundary conditions to be satisfied by the electric  field solutions 
corresponding to the quasinormal modes are:
\begin{eqnarray}
E(0)  &=&  0 \,, \\
E(z_h) &=&  e^{-i\omega r_*(z_h) + \frac{\zeta(z_h)}{2}} \,.
\end{eqnarray} 
Equations of motion  (\ref{eqT})  and (\ref{eqL})  do not present analytic solutions, so they are solved numerically. The more convenient way to perform the numerical computations is to  impose boundary conditions for the electric field and for the electric field derivative at the horizon: 
\begin{eqnarray}
E(z_h) &=&  e^{-i\omega r_*(z_h) + \frac{\zeta(z_h)}{2}} \,, \\
E^{'}(z_h)  &=&  \left(-i\omega r'_{*}(z_h) + \frac{\zeta'(z_h)}{2}+c^{(1)}\right) E(z_h) \,.
\label{BCHorizon}
\end{eqnarray} 
Then one searches  for the lowest complex frequency that lead to  a solution vanishing at $ z = 0$. 
These type of solutions, that depend on the temperature and on the magnetic field,   are the ones that  represent the charmonium quasistates. Then, one uses these solutions in the calculation of the energy density in eqs. (\ref{rhoT}) and (\ref{rhoL}). 
 
The fields, and consequently the density, depend only  on the variable $z$. So, we need just the    
  Fourier transform of  $ \rho (z)$ in coordinate $z$:  ${\tilde \rho (k)}  $. It is convenient,  for the computation of the CE, in this one dimensional case,  to  write ${\tilde \rho} (k) = \left( C(k) -  iS(k)\right) / \sqrt{2\pi} $, where
\begin{eqnarray}
\label{Fourier1}
C(k)&=&\int_{0}^{z_h}\rho(z)\cos({kz})dz \,,\\
\label{Fourier2}
S(k)&=&\int_{0}^{z_h}\rho(z)\sin({kz})dz \,.
\end{eqnarray}
In terms of these components,  the  modal fraction reads:
\begin{equation}
\epsilon (k)=\frac{S^2(k)+C^2(k)}{ \left[ S^2(k)+C^2(k) \right]_{max}}\,,
\label{1DModalfrac}
\end{equation}
and the  DCE (\ref{momentumentropy})  takes the form:
\begin{equation}\label{CE}
 S  = -\int^{\infty}_{-\infty}   \epsilon (k)\log{ \left[  \epsilon (k)\right]  } \,dk\,.
\end{equation}

\section{Results}

\begin{figure}[t]
\centering
\includegraphics[scale=0.30]{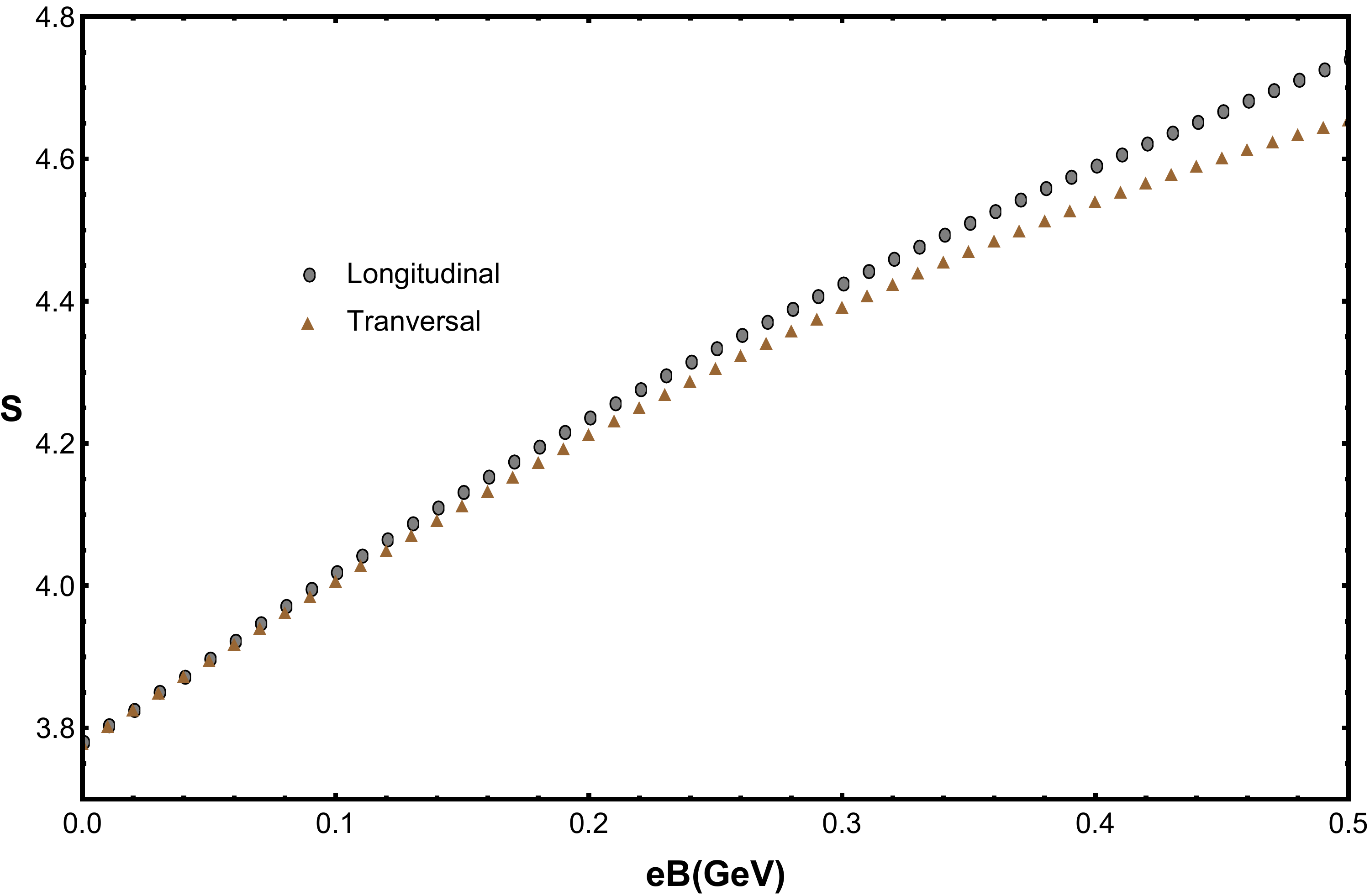}
\caption{Differential configuration entropy $ S$ for charmonium   J/$\psi$  as a function of the  magnetic field $eB$  at  temperature  $T \to 0 $ MeV}
\label{temp0}
\end{figure}

\begin{figure}[t]
\centering
\includegraphics[scale=0.30]{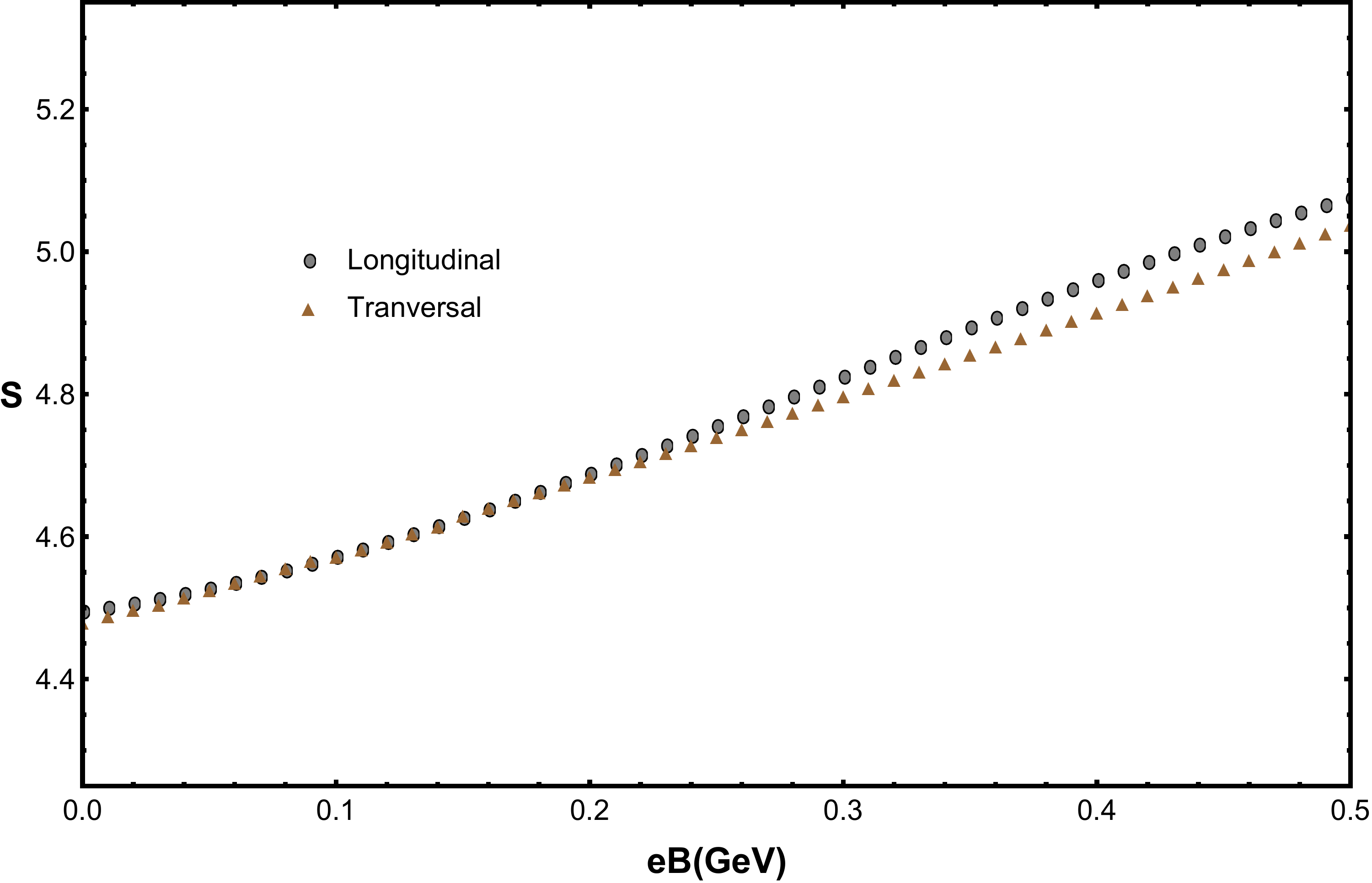}
\caption{ Differential configuration entropy $ S$ for charmonium   J/$\psi$  as a function of the magnetic field $eB$ at  temperature   $T$   = 100  MeV}
\label{temp100}
\end{figure}

 The strategy for studying the dependence of the charmonium DCE with the magnetic field is the following. We calculate the fields $ E_3$  and $E_\alpha $ ($\alpha = 1,2$) ,  with complex frequencies,  that solve equations (\ref{eqT})  and (\ref{eqL}) and have the asymptotic form of eq. (\ref{BCHorizon})  and vanish at $ z= 0$. Then we insert  these solutions into the expressions for  the energy densities  (\ref{rhoL}), in the longitudinal case,  or   (\ref{rhoT}) in the transversal case. 
Finally, the DCE is obtained using, in this order,  equations  (\ref{Fourier1}),    (\ref{Fourier2}),  (\ref{1DModalfrac}) and (\ref{CE}). 

\begin{figure}[t]
\centering
\includegraphics[scale=0.30]{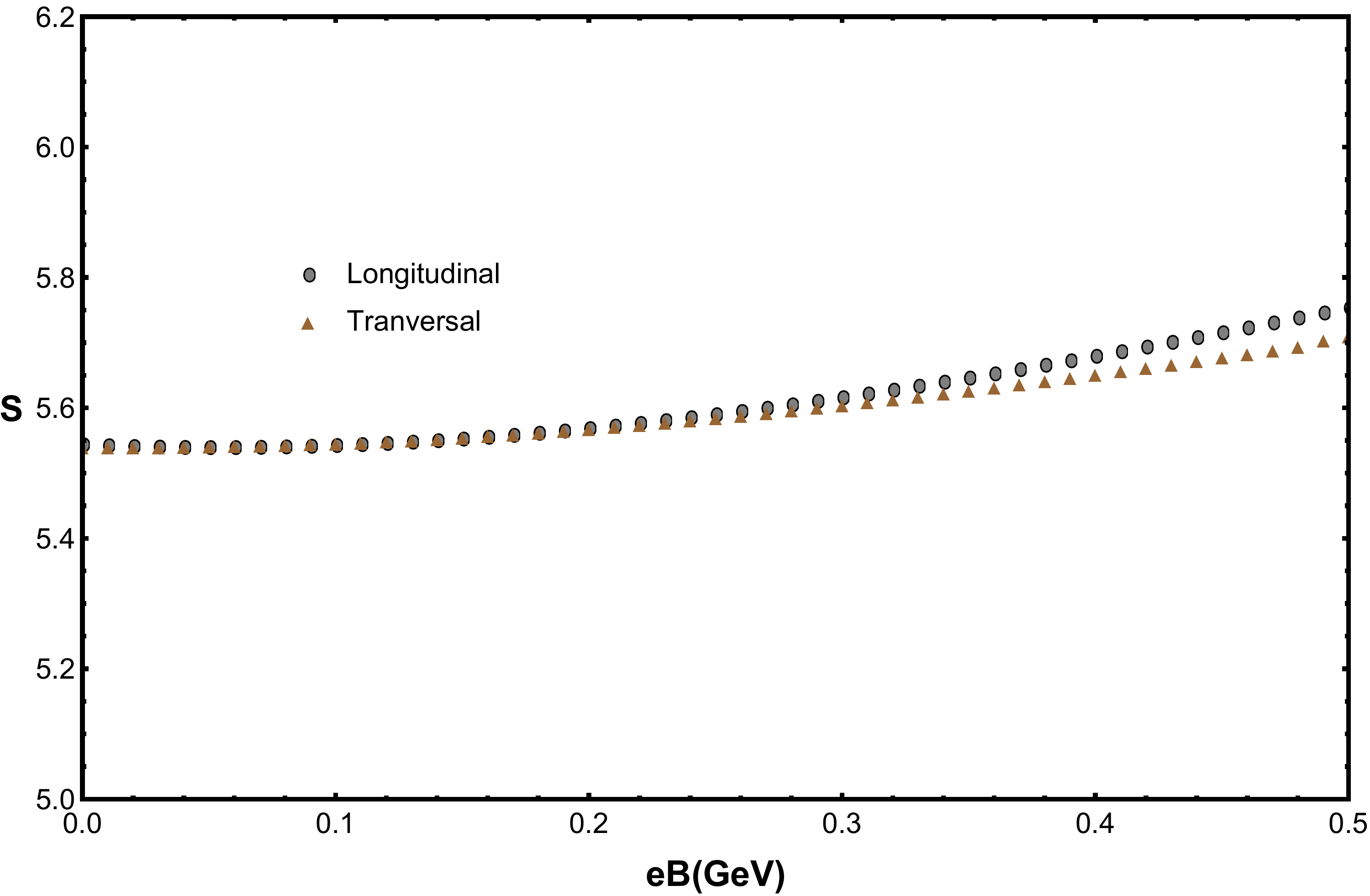}
\caption{ Differential configuration entropy $ S$ for charmonium   J/$\psi$  as a function of the magnetic field $eB$ at  temperature   $T$   = 200  MeV.}
\label{temp200}
\end{figure}

\begin{figure}[t]
\centering
\includegraphics[scale=0.30]{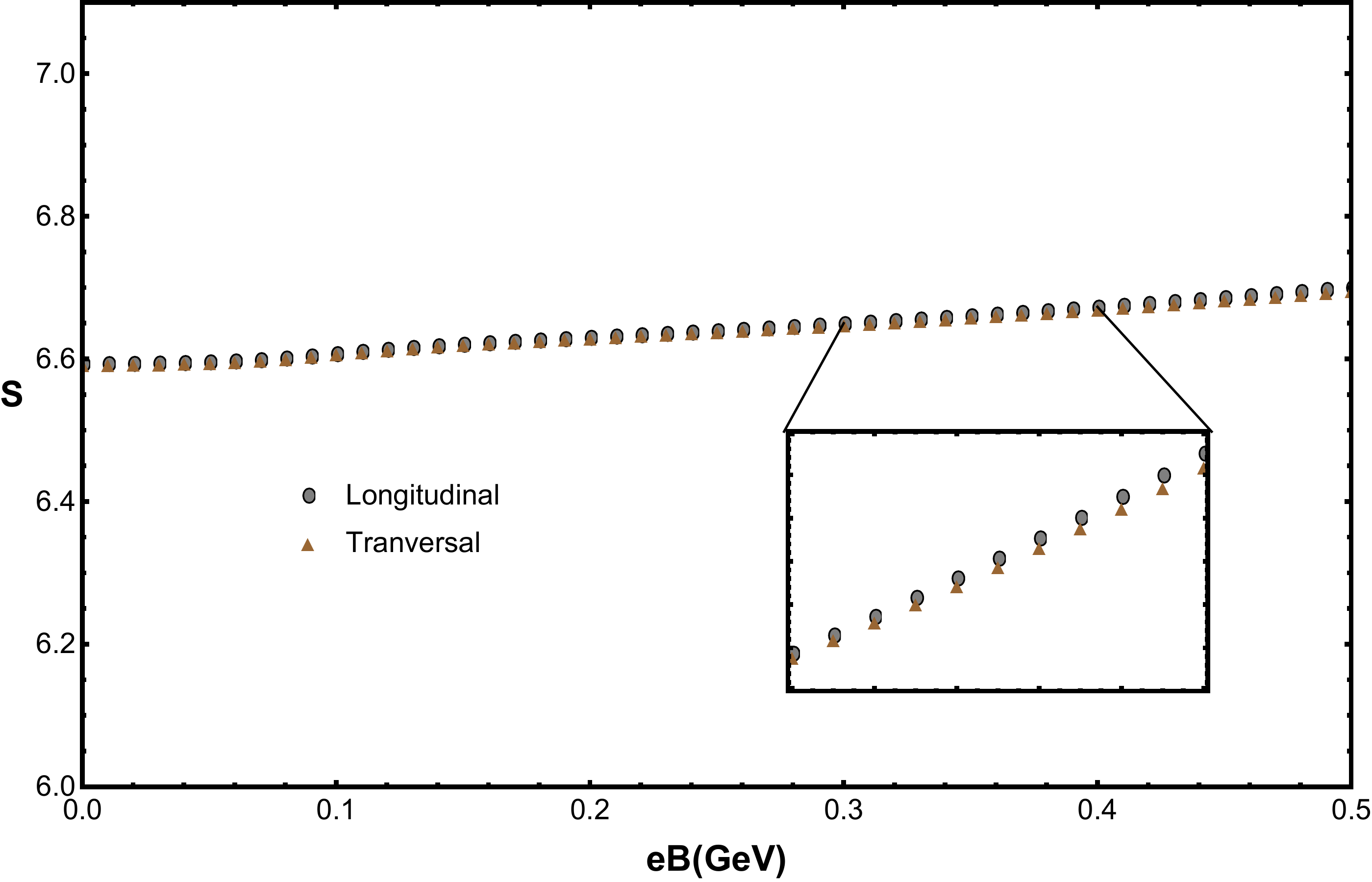}
\caption{
Differential configuration entropy $ S$ for charmonium   J/$\psi$  as a function of the magnetic field $eB$ at  temperature   $T$   = 300  MeV, with a x=zoom of the region 0.3 $ \le eB \le 0.4 $ by a factor of 5.  }
\label{temp300}
\end{figure}

We show in figures   \ref{temp0}, \ref{temp100}, \ref{temp200} and  \ref{temp300}     the DCE for charmonium  J/$\psi$ as a function of the magnetic field $eB$  when the plasma is at temperatures of, respectively,  $ T = 0, 100, 200, 300$ MeV, for both longitudinal and transversal polarisations. One notes that the DCE increases with the field $eB$ and this effect is more intense for lower temperatures and very similar for the two polarisations. These figures are plotted using the same scale in order to provide a comparison of the DCE variations. In order to make it possible to see that even for the higher temperatures the DCE increases  with temperature, we inserted inside  figure \ref{temp300}  a small plot with a zoom in  the region 0.3 GeV $  \le eB \le $ 0.4 GeV. 
The scale was amplified in this part  by a factor of 5  so that one can see that indeed the DCE is increasing.

\begin{figure}[t]
\centering
\includegraphics[scale=0.30]{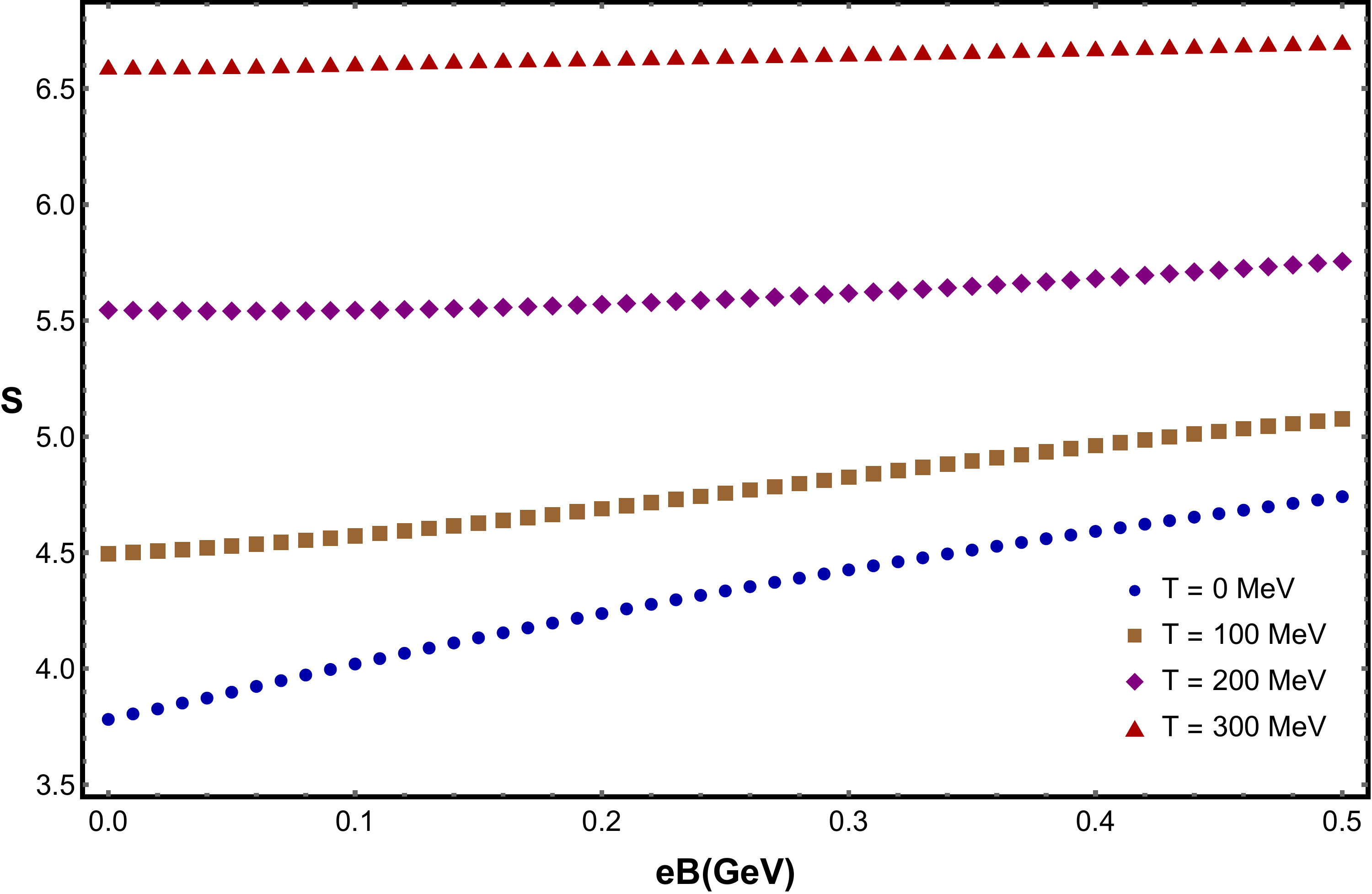}
\caption{
Differential configuration entropy $ S$ for charmonium   J/$\psi$  as a function of the magnetic field $eB$,  in the longitudinal polarization case, at  temperatures   $T$   = 0, 100, 200, 300  MeV.  }
\label{longpolarization}
\end{figure}

\begin{figure}[t]
\centering
\includegraphics[scale=0.30]{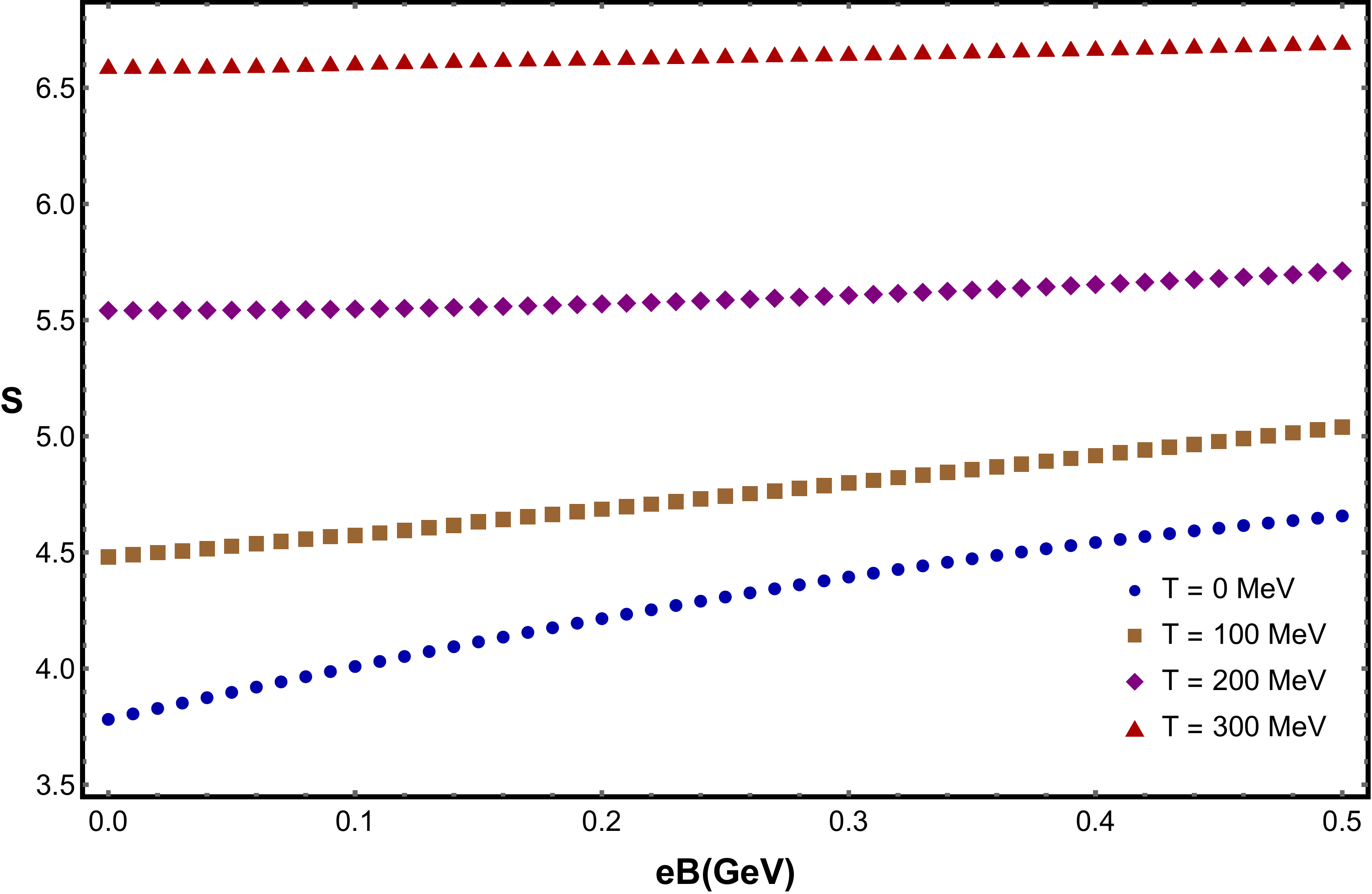}
\caption{
Differential configuration entropy $ S$ for charmonium   J/$\psi$  as a function of the magnetic field $eB$, in the transversal polarization case,  at  temperatures   $T$   = 0, 100, 200, 300  MeV.     }
\label{tranvpolarization}
\end{figure}

Then, in order to make it clear the effect of the temperature, we plot in figure 
\ref{longpolarization} the results for the four temperatures considered in the previous figures for the longitudinal polarisation case. Then, the same thing is shown in figure \ref{tranvpolarization}  but  for transversal polarisation. One notices that there is a clear increase of the DCE with temperature. This is consistent with the fact that as the temperature increases the charmonium state becomes more unstable against dissociation in the thermal medium.

 Now, let us see  if  the dependence of the DCE on the magnetic field $ e B$ can be  approximated by  some  simple   form.  Searching for polynomial approximations, one finds a very nice fit using a second order expression of the form:
 \begin{equation}\label{scalinglaw}
 S = c_0  + c_1 (eB) + c_2 (eB)^{2},
\end{equation}
\noindent with $c_0, c_1 , c_2$ varying with the temperature. Tables { \bf 1}  and  {\bf 2}   show the values of the parameters at the temperatures considered previously and the mean absolute error $R^{2}_{Adj}$ for the  approximation of the DCE of    J/$\psi$ by relation (\ref{scalinglaw}).

\begin{table}[h]\label{data1}
\centering
\begin{tabular}{|l||l||l||l||l|}
\hline
\multicolumn{5}{|c|}{Coefficients for J/$\psi$ longitudinal polarisation DCE fit }\\ \hline \hline
T (GeV)&    $ c_0$&$ c_1 (GeV)^{-1}$ & $  c_2(GeV)^{-2}$ & Mean absolute error\\ \hline
0 & 3.7788 $\pm$0.0009   & 2.528 $\pm$0.008      & -1.23 $\pm$0.02     &  0.0018 \\ \hline
0.1 & 4.476 $\pm$0.003 & 1.00$\pm$0.03   & 0.48 $\pm$0.02   & 0.007  \\ \hline
0.2 & 5.5399 $\pm$0.0009 & -0.028 $\pm$0.009  & 0.94 $\pm$ 0.02   & 0.0019 \\ \hline
0.3 & 6.5895 $\pm$0.0007 & 0.179 $\pm$0.006    & 0.07 $\pm$0.01   &  0.0014 \\ \hline
\end{tabular}
\caption {Coefficients $ c_0$, $ c_1$ and $ c_2$ of  eq. \ref{scalinglaw} for longitudinal polarisation at different temperatures.}
\end{table}

\begin{table}[h]\label{data2}
\centering
\begin{tabular}{|l||l||l||l||l|}
\hline
\multicolumn{5}{|c|}{Coefficients for J/$\psi$ transversal polarisation DCE fit }\\ \hline \hline
T (GeV)&   \centering $ c_0 $&$  c_1 (GeV)^{-1}$ & $  c_2(GeV)^{-2}$ & Mean absolute error\\ \hline
0 & 3.7771 $\pm$0.0007   & 2.4806 $\pm$0.007    & -1.43 $\pm$0.01   &  0.0015 \\ \hline
0.1 & 4.477 $\pm$0.0007   & 0.983 $\pm$0.007     & 0.29 $\pm$0.01    & 0.0011  \\ \hline
0.2 & 5.5393 $\pm$0.0005   & 0.033 $\pm$0.005     & 0.625 $\pm$0.009    & 0.0011  \\ \hline
0.3 & 6.5895 $\pm$0.0007 & 0.1863 $\pm$0.007  & 0.05 $\pm$0.01  & 0.0014 \\ \hline
\end{tabular}
\caption {Coefficients $ c_0$, $ c_1$ and $ c_2$ of  eq. \ref{scalinglaw} for transversal polarisation at different temperatures.}
\end{table}

\begin{figure}[t]
\centering
\includegraphics[scale=0.40]{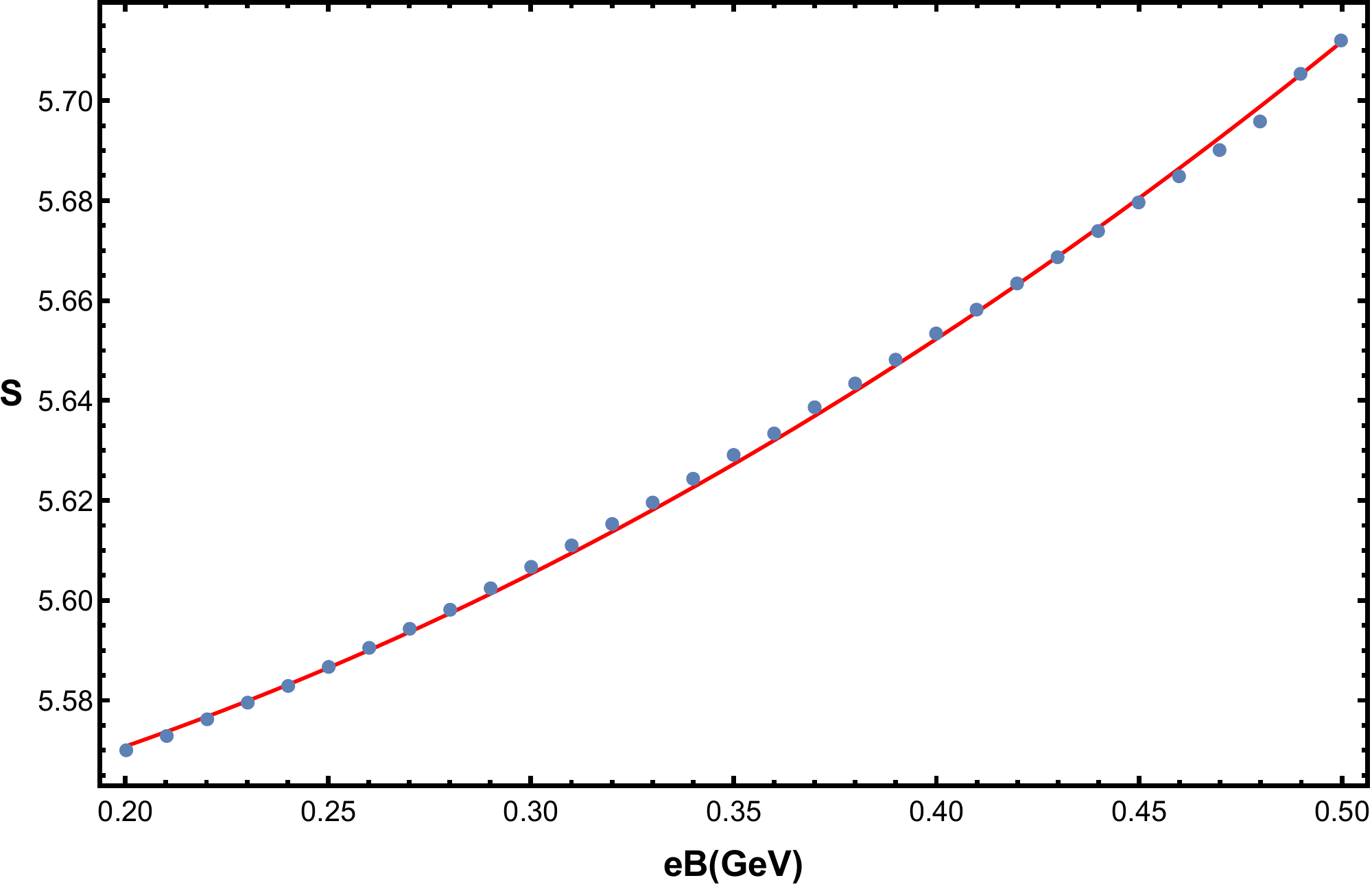}
\caption{ Solid red line: quadratic fit obtained using expression  (\ref {scalinglaw}) with the coefficients given on table 2. Dots:  actual values of the DCE at temperature   T   =  200 MeV.     }
\label{scalingT200transversepolarization}
\end{figure}

For the sake of illustrating  the quality of the fit of the DCE by the quadratic polynomials of the form given in eq. (\ref {scalinglaw}), we show in figure \ref{scalingT200transversepolarization} the case of transverse polarisation at T = 200 MeV.

\section{Conclusions}
 It is known that the presence of $eB$ magnetic fields enhances the    dissociation effect of charmonium in a plasma\cite{Braga:2019yeh}. The higher  is the field, the  strongest is the effect. The   thermal 
 dissociation corresponds to the disappearance of the charmonium quasistates in the medium. 
 So, an increase in the dissociation intensity corresponds to an increase in the instability  of charmonium. As discussed in the introduction, for many different systems it was observed that the 
 configuration entropy  works an an indicator of stability. The more stable is the system, the lower is the value of the configuration entropy. In this article we have calculated   the DCE of charmonium quasistates in a plasma in the presence of magnetic fields and obtained the result that it increases with the $eB$ field. This is consistent with the increase in instabily associate with dissociation in the medium.
 We also found that the DCE increases with the temperature, as it was previously observed in \cite{Braga:2020myi},   and is also consistent  with the increase in instability caused by the enhancement of the dissociation effect with the temperature. 
 
 A result that also  emerged here is that the variation of the DCE with the magnetic field is more intense for lower temperatures. This  is consistent  with the fact observed in ref.\cite{Braga:2019yeh}   that the increase in the imaginary part of the quasinormal mode frequencies caused by the magnetic field is  larger for lower temperatures. The imaginary part of the complex frequencies represents the thermal width, that is related to the dissociation level. The larger it is, the more unstable is the quasistate. 
 
 It is important to remark that here we  considered only the   effect of the magnetic field  in the thermal medium. It is important to note that it is possible to study the direct effect of the field on the charged constituents of the meson as was considered in  \cite{Dudal:2014jfa,Dudal:2015kza,Dudal:2018rki}.

\noindent {\bf Acknowledgments:}  The authors are supported by  CNPq - Conselho Nacional de Desenvolvimento Cientifico e Tecnologico (N.B by grant  307641/2015-5 and R. M. by a graduate fellowship). This work received also support from  Coordenação de Aperfeiçoamento de Pessoal de Nível Superior - Brasil (CAPES) - Finance Code 001.

 \end{document}